\documentstyle[12pt]{article}
\newcommand{\R}{\mathop{\rm Re}\nolimits}
\newcommand{\I}{\mathop{\rm Im}\nolimits}
\newcommand{\sgn}{\mathop{\rm sgn}\nolimits}
\newcommand{\eps}{\varepsilon}
\newcommand{\dsp}{\displaystyle}

\newcommand{\Eg}{{\cal E}_G^{\scriptscriptstyle(1)}{}}
\newcommand{\Eq}{{\cal E}_Q^{\scriptscriptstyle(1)}{}}
\newcommand{\Egh}{{\cal E}_{\rm gh}^{\scriptscriptstyle(1)}{}}
\newcommand{\F}{{}^*{\!F}}
\newcommand{\eff}[1]{_{\rm eff\,#1}}
\newcommand{\p}{{\bf p}}
\newcommand{\tth}{\tilde\theta}
\newcommand{\fr}[2]{{\textstyle\frac{#1}{#2}}}

\begin{document}
\begin{titlepage}
\begin{center}
\large\bf One-loop Energy Corrections in the Nonabelian Gauge Field Theory in
2+1 Dimensions
\end{center}
\vspace{0.5cm}
\renewcommand{\thefootnote}{\fnsymbol{footnote}}
\begin{center}
V.~Ch.~Zhukovsky, N.~A.~Peskov\footnote{E-mail addresses:
{\tt th180@phys.msu.su} and {\tt peskov\_\,nick@mail.ru}.}\\
{\sl Physical Faculty, Moscow State University,\\
119899, Moscow, Russia}
\end{center}
\begin{abstract}
Exact constant solutions of field equations in the classical nonabelian
$SU(2)$ field theory with the Chern-Simons topological mass in
$2+1$-dimensional space-time have been obtained. One-loop contributions of
boson and fermion fluctuations to the gauge field energy have been calculated.
\end{abstract}
\vspace{0.3cm}
\vfill
\end{titlepage}
It is well known that a topological term, the so-called Chern-Simons term,
which plays the role of the gauge field mass, can be introduced in
odd-dimensional gauge theories~\cite{CSterm,Jackiw}.

As an example of such theories, we consider the three-dimensional
topologically massive $SU(2)$-gluodynamics described by the
Lagrangian~\cite{CSterm}
\begin{equation}
{\cal L} = -\frac14F_{\mu\nu}^a F^{a\,\mu\nu}-\frac\theta4
\eps^{\mu\nu\alpha}\left(F_{\mu\nu}^a A_{\alpha}^a-\frac{g}3
\eps^{abc}A_{\mu}^a A_{\nu}^b A_{\alpha}^c\right).
\label{LagrQCD}
\end{equation}
Here $\mu,\nu = 0,1,2$; $a = 1,2,3$, potentials $A_{\mu}\equiv\tau^a
A_{\mu}^a/2$, where $\tau^a$ are Pauli matrices in color space, the color
field tensor is given by $F_{\mu\nu}^a = \partial_\mu^{}A_{\nu}^a-
\partial_\nu^{}A_{\mu}^a+g\eps^{abc}A_{\mu}^bA_{\nu}^c$. The coefficient
$\theta$ in front of the last term in (\ref{LagrQCD}) (Chern-Simons term) is
the Chern-Simons (CS) mass of the gauge field.

In this case the field equations take the form
\begin{equation}
D_{\mu}^{ac}F^{a\,\mu\beta}-\frac\theta2\eps^{\beta\mu\nu}F_{\mu\nu}^c = 0,
\label{YMeq}
\end{equation}
where $D_{\mu}^{ac}=\delta^{ac}\partial_{\mu}-g\eps^{abc}A_{\mu}^b$. It is
easily verified that these equations are satisfied by the following constant
nonabelian potentials
\begin{equation}
A^{a\,\mu} = \frac{\theta}{2g}\delta^{a\mu}\chi_{\lambda\omega}^{(a)},
\label{field}
\end{equation}
with normalized constant vector $\chi_{\lambda\omega}^{(a)} = (\lambda i,\,
\lambda\omega i,\,\omega)$, where $\lambda = \pm1$ and $\omega = \pm1$. Note
that $\lambda$ and $\omega$ take its values independently. The Kronecker delta
$\delta^{a\mu}$ in (\ref{field}) implies that directions 1, 2, 3 in the color
space correspond to directions 1, 2, 0 in the Minkowski $2+1$ space-time,
respectively. The following relations are easily verified:
$$
F_{\mu\nu}^a = -\frac\theta2\eps_{\mu\nu\alpha}A^{a\,\alpha},\qquad
A^{a\,\mu}A_{\mu}^b = \frac{\theta^2}{4g^2}\delta^{ab},\qquad
A^{a\,\mu}A^{a\,\nu} = \frac{\theta^2}{4g^2}g^{\mu\nu}.
$$

The gauge field energy density $E$ corresponds to the ${\cal T}^{00}$
component of the symmetric energy-momentum tensor ${\cal T}^{\mu\nu}$ obtained
by the standard method~\cite{STZB}
$$
{\cal T}^{\mu\nu} = T^{\mu\nu}+\partial_{\alpha}T^{\alpha\mu\nu} =
\F^{a\,\mu}\F^{a\,\nu}-\frac12g^{\mu\nu}\F_{\alpha}^a\F^{a\,\alpha}.
$$
Here $T^{\mu\nu}$ is the canonical energy-momentum tensor, $T_{\alpha\mu\nu} =
-A_{\mu}^a\left(F_{\alpha\nu}^a+\frac12\theta\eps_{\alpha\nu\beta}A^{a\,\beta}
\right)$, and $\F_{\mu}^a = \frac12\eps_{\mu\alpha\beta}F^{a\,\alpha\beta}$,
the dual field tensor, is in fact a vector in $2+1$-dimensional space-time.
Taking the solution (\ref{field}) into account we can now represent
${\cal T}^{\mu\nu}$ in the form
$$
{\cal T}^{\mu\nu} = -\frac{\theta^4}{32g^2}g^{\mu\nu}.
$$
The negative sign of the energy density $E = {\cal T}^{00}$ means that, for
the given value of $\theta\ne0$, the energy of this solution is lower than
that of the trivial vacuum solution $A^{\mu} = 0,$ obtained in the
perturbation theory.

Next, we consider the one-loop effective potential
$$
V\eff{} = E+\Eg+\Eq
$$
with consideration for both gluon $\Eg$ and quark $\Eq$ loops. We start with
the gluon contribution $\Eg$. The total gauge field ${\cal A}^{a\,\mu}$ is
represented as the sum of the constant classical background field
(\ref{field}) and quantum fluctuations $a^{a\,\mu}$, i.e.
$$
{\cal A}^{a\,\mu} = A^{a\,\mu}+a^{a\,\mu}.
$$

Introducing the gauge-fixing term
$$
{\cal L}_{\rm gf} = -\frac1{2\xi}\left(\nabla_\mu^{ab}a^{b\,\mu}\right)^2,
$$
where $\nabla_\mu^{ab} = \delta^{ab}\partial_\mu+g\eps^{abc}A^c_\mu$ is the
background covariant derivative, we obtain the fluctuation Lagrangian for the
gauge $\xi = 1$ in the form
$$
{\cal L}(a_{\mu}^b|A_{\mu}^b) = {\cal L}(a_{\mu}^b)-
\frac12(\partial_\mu a^{b\,\mu})^2-\frac{\theta^2}4(a^{b\,\mu})^2+
g\eps^{abc}A_\mu^a f^{b\,\mu\nu} a_\nu^c-
g\eps^{abc}A_\mu^c a^{b\,\mu} \partial_\nu a^{a\,\nu}.
$$
Here, the first term in r.h.s.\ is the same Lagrangian as in (\ref{LagrQCD}),
though defined in terms of the field fluctuations $a_{\mu}^b$ and the
corresponding tensor $f_{\mu\nu}^b$.

The energy spectrum of gluon fluctuation modes $\eps_i$, corresponding to
{\it all} degrees of freedom (physical and unphysical) that remain after gauge
fixing, is determined by the equation
$$
\det\left[\delta^{ac}g^{\mu\nu}\left(p^2+\fr12\theta^2\right)-
i\theta\delta^{ac}\eps^{\alpha\mu\nu}p_\alpha-
2ig\eps^{abc}\left(A^{b\,\nu}p^\mu+A^{b\,\mu}p^\nu-
g^{\mu\nu}A_\alpha^b p^\alpha\right)\right] = 0.
$$
Nine solutions of the above equation are as follows:
$$
\eps_i^2 = \p^2+\lambda_i\frac{\theta^2}2,
$$
where $\lambda_{1,2} = 3\pm2\sqrt2$, $\lambda_{3,4} = 5\pm2\sqrt6$, and the
other five degenerate spectrum branches with $\lambda_{5,\dots,9} = -1$,
corresponding to tachyonic modes, are unstable.

The one-loop energy correction due to gauge field fluctuations around the
background is given by the expression
\begin{equation}
\Eg = \fr12\sum_{\p,i}\eps_i(\p) =
\fr12\sum_{\p,i}\,[\p^2+\lambda_i\theta^2/2]^{1/2}.
\label{BosonE}
\end{equation}
Consideration for tachyonic modes contribution to expression (\ref{BosonE})
results in the appearance of the imaginary part of the energy correction
$\Eg$, defined as follows (see, e.g.,~\cite{STZB}):
$$
\I\Eg = -\fr12\sum_{\eps_i^2<0}|\eps_i^2(\p)|^{1/2}.
$$
The imaginary part is finite and its calculation results in the following
expression (with due regard for degeneracy)
$$
\I\Eg = -\frac5{2(2\pi)^2}\int_{|\p|=0}^{\theta/\sqrt2}
\sqrt{\frac{\theta^2}2-\p^2}\,d\p = -\frac5{24\sqrt2\pi}|\theta|^3.
$$

The real part of the energy correction corresponding to $\eps_i^2>0$
contribution to (\ref{BosonE}) is divergent. The appropriate application of
the dimensional regularization procedure yields:
\begin{equation}
\R\Eg^{reg} = \fr12\mu^{2\epsilon}\sum_{\eps_i^2>0}\,
[\p^2+\lambda_i\theta^2/2]^{1/2-\epsilon},
\label{Re_E_reg}
\end{equation}
where $\mu$ is an arbitrary scale mass parameter. We can now introduce the
proper time representation for this expression, and using the regulator
$\epsilon>\frac12$, we find
$$
\R\Eg^{reg} = \frac{\mu^{2\epsilon}}{2(2\pi)^2}
\frac1{\Gamma(\epsilon-1/2)}\int_0^\infty\frac{ds}{s^{3/2-\epsilon}}
\biggl\{\int\!d\p\sum_{i=1}^4e^{-s(\p^2+\lambda_i\theta^2/2)}+
5\int_{|\p|=\theta/\sqrt2}^\infty d\p\,e^{-s(\p^2-\theta^2/2)}\biggr\},
$$
and this, upon integration over the two-dimensional momentum, yields
$$
\R\Eg^{reg} = \frac{\mu^{2\epsilon}}{8\pi\Gamma(\epsilon-1/2)}
\int_0^\infty\frac{ds}{s^{5/2-\epsilon}}
\biggl\{5+\sum_{i=1}^4e^{-s\lambda_i\theta^2/2}\biggr\}.
$$
The (ultraviolet) divergent at $s\to0$ term in $\R\Eg^{reg}$ is determined by
the expression
\begin{equation}
\R\Eg^{div} = \frac{\mu^{2\epsilon}}{8\pi\Gamma(\epsilon-1/2)}
\int_0^\infty\frac{ds}{s^{5/2-\epsilon}}
\biggl\{5+\sum_{i=1}^4\Bigl(1-s\lambda_i\frac{\theta^2}2\Bigl)\biggr\}.
\label{E_div}
\end{equation}
The difference $\R\Eg^{ren}\equiv\R\Eg^{reg}-\R\Eg^{div}$ is finite in the
limit $\epsilon\to0$ and its calculation can be reduced to $\Gamma$-function
integrals. However, the two results, for imaginary and real parts, still
include contributions of the unphysical modes introduced by the gauge-fixing
term. In order to get rid of these unphysical contributions, the ghost fields
$\eta^a$ and $\overline\eta^{\,a}$ with the Lagrangian
$$
{\cal L}_{\rm gh} = (\nabla^{ab}_\mu\overline\eta^{\,b})(D^{ac\,\mu}\eta^c)
$$
should be introduced. Here $D_\mu^{ab} = \delta^{ab}\partial_\mu+
g\eps^{abc}{\cal A}^c_\mu$ is the total covariant derivative. The ghost energy
spectrum is determined by the equation:
$$
\det(\nabla^2)^{ab} = 0,
$$
where $(\nabla^2)^{ab} = \nabla^{ac}_\mu\nabla^{cb\,\mu}$. Thus, three
branches of the ghost spectrum are obtained $\eps^2_{{\rm gh}\,i} = \p^2+
\frac12\zeta_i\theta^2$, where $\zeta_1 = -1$, $\zeta_{2,3} = \pm i$, and the
ghost contribution to the effective potential is given by the general formula:
$$
\Egh = -\sum_{\p,i}\,\eps_{{\rm gh}\,i}(\p).
$$
Here, the ``minus'' sign arises as a result of the Fermi statistics applied to
ghosts. The branch $\eps_{{\rm gh}\,1}$ makes a nonzero contribution only to
the imaginary part of effective potential. This is easily demonstrated in
calculations like those for the gluon contribution to the energy imaginary
part. The total result is the following
$$
\I\Eg+\I\Egh = -\frac7{24\sqrt2\pi}|\theta|^3.
$$
It is also easy to see that the other energy branches make contribution to
the real part only
$$
\R\Egh^{reg} = -\mu^{2\epsilon}\sum_{\p,i=2,3}
[\p^2+\fr12\zeta_i\theta^2]^{1/2-\epsilon} =
-\frac{\mu^{2\epsilon}}{2\pi\Gamma(\epsilon-\frac12)}\int_0^\infty
\frac{ds}{s^{5/2-\epsilon}}\cos\frac{s\theta^2}2.
$$
Finally, we obtain the following result for the boson contribution to the
renormalized effective potential, where only the gluon {\it physical\/}
degrees of freedom have been retained
$$
V\eff{}\equiv E+\R\Eg+\R\Egh = -\frac{\theta^4}{32g^2}-
\frac{|\theta|^3}{2\pi}\biggl(1+\frac{3\sqrt6}4\biggr).
$$

We observe in this connection that the standard prescription for the
dimensional regularization integrals, similar to (\ref{E_div}), implies that
they are assumed to be equal to zero~\cite{ren}. The terms like $1/\epsilon$
evidently do not appear in the divergent parts. Hence, one may say that the
divergent part of the effective potential is put equal to zero by definition,
i.e., $\R\Eg^{div} = 0$ and $\R\Egh^{div} = 0$, and moreover, masses and
coupling constants are not renormalized in this case, which is in agreement
with the fact that gauge theories are super-renormalizable in three
dimensions~\cite{Jackiw,Sren}.

It should also be reminded that the tachyon modes of the gluon spectrum make
no contribution to the real part of the correction~(\ref{BosonE}).

We now pass to calculation of the fermion contribution to the effective
potential. The quark energy spectrum in the external field specified by
(\ref{field}) can be found from the Dirac equation
\begin{equation}
\big[(i\partial_\mu\gamma^\mu-m)+
\fr12g\tau^a A_\mu^a\gamma^\mu\big]\psi(x) = 0.
\label{Dirac}
\end{equation}
Here, the following representation for the $\gamma$-matrices in $2+1$
dimensions is chosen: $\gamma^0 = \sigma^3$, $\gamma^{1,2} = i\sigma^{1,2}$,
with $\sigma^i$ as Pauli matrices. The $\gamma$-matrices obey the following
relation: $\gamma^\mu\gamma^\nu = g^{\mu\nu}-i\eps^{\mu\nu\alpha}
\gamma_\alpha$.

Two quantum states of quark in the $SU(2)$ fundamental representation
correspond to two different colors. Spinors in $2+1$-dimensions are
two-component objects, which corresponds to two degrees of freedom of a
fermion (particle--antiparticle) and, hence, spin coefficients are closely
related to the sign of the particle energy. The mass term $m{\overline\psi}
\psi$ in the Lagrangian is not invariant under $\cal P$-inversion of
coordinates: $x\to-x$, $y\to y$, or $x\to x$, $y\to- y$. Recall that
simultaneous reflection of two spatial coordinates is equivalent to rotation.
The Lagrangian is invariant only under the combined operation that involves
$\cal P$-inversion and the substitution $m\to-m$~\cite{CSterm,Spin}.
Therefore, the spin operator for two-dimensional fermions is naturally
introduced in the form $\Sigma=(\sgn m)\gamma^0/2$. Since the matrix
$\gamma^0$ simultaneously determines the sign of the energy, the positive- and
negative-energy states correspond to opposite signes of the spin polarization.

Solving equation (\ref{Dirac}) in the gauge field (\ref{field}) results in two
different branches of the quark energy spectrum
$$
\eps_1^2=\p^2+m\eff1^2,\qquad\eps_2^2=\p^2+m\eff2^2,
$$
where $m\eff1^2 = (m-\tth)^2$, $m\eff2^2 = (m-\tth)(m+3\tth)$ and $\tth =
\theta/4$. It is interesting to note that in the special case, when $m =
\tth$, the quark effective mass in this gauge field vanishes, and the two
branches of the energy spectrum coincide. For values of the mass lying in the
interval $-\tth-2|\tth|<m<-\tth+2|\tth|$, the energy squared, $\eps_2^2$,
becomes negative for certain values of the quark momentum $\p$, and tachyonic
modes appear in the quark spectrum.

Two branches of the energy spectrum are related to opposite projections of the
particle color spin corresponding to two eigenvalues of the operator
\begin{equation}
{\bf J} = J^a\tau^a/2 = \tth^{-1}gA^\mu p_\mu-
\omega\gamma^0\tau^3(p_\mu\gamma^\mu+\tth-m)
\label{J}
\end{equation}
that is defined for the plane-wave solutions of the Dirac equation
(\ref{Dirac}) $\psi_s(x) = \exp(-i\eps_s t+i\vec{px})\Psi_s$ with $s = 1,2$:
$$
{\bf J}\Psi_{s} = (-1)^s(\tth-m)\Psi_{s},
$$
where $\Psi_s$ are constant spinors, corresponding to the Dirac Hamiltonian
eigenvalues $\eps_s$. We note, that the r.h.s.\ of the above equation vanishes
in the case $m = \tth$. Moreover, under this condition, it appears that
$\Psi_1 = \Psi_2$.

The one-loop quark contribution to the vacuum energy is described by the
general equation
\begin{equation}
\Eq = \R\Eq+i\I\Eq = -\sum_{\eps_s^2(\p)>0}\eps_s(\p)-
i\sum_{\eps_s^2(\p)<0}|\eps_s(\p)|.
\label{FermionE}
\end{equation}
The real part of the correction is defined by the expression
$$
\R\Eq = -\frac1{(2\pi)^2}\int\!d\p\,\eps_1-
\frac1{(2\pi)^2}\int_{|\p|=p_*}^{\infty}d\p\,\eps_2.
$$
Here $p_* = [-m\eff2^2]^{1/2}\Theta(-m\eff2^2)$ is the lowest of the quark
momentum values that provide a real contribution to the quark energy in the
absence of the tachyonic modes. Function $\Theta(x)$ is the Heaviside (unit
step) function. To find the finite part of the quark energy correction, the
same procedure as in the case of gluons can be applied. The negative sign in
front of the real part of the energy in (\ref{FermionE}) is due to the fact
that quarks are fermions. The quark contribution to the effective potential
finally reads
$$
\begin{array}{c}
\R\Eq = \dsp{
\frac1{6\pi}\Big\{|m-\tth|\Big((m-\tth)^2-\frac{3|m|}4|m+3\tth|\Big)-
\frac{|m|}4(m-\tth)(5m-\tth)+}\\[2.2ex]
|m|\tth^2+[(m-\tth)(m+3\tth)]^{3/2}\Theta((m+\tth)^2-4\tth^2)\Big\}.
\end{array}
$$
We note that the sign of this expression is opposite to that of the gluon
contribution.

Like in the gluon case, the imaginary part of the energy does not diverge and
takes nonzero values only if tachyonic modes are present. The final expression
for the imaginary part can be written in the following simple form
$$
\I\Eq = -\frac1{6\pi}|(m-\tth)(m+3\tth)|^{3/2}\Theta(4\tth^2-(m+\tth)^2).
$$

In particular, in the limit of the vanishing quark mass, $m\to0$, we obtain
$$
\R\Eq = \frac{|\tth|^3}{6\pi},\qquad\I\Eq = -\frac{\sqrt3|\tth|^3}{2\pi}.
$$

It should be emphasized that it is the tachyonic modes in the quark spectrum
that lead to the appearance of an imaginary part in the energy of vacuum
fluctuations. This means that the constant gauge field configuration in
question is unstable. However, such factors, as boundaries and finite volume
of the spatial region occupied by the gauge field, finite temperature, as well
as consideration for inhomogeneous modes and higher loop contributions may
lead to its stabilization.


\begin{thebibliography}{99}
\bibitem{CSterm}
{\it Deser~S., Jackiw~R., Templeton~S.}
Ann. Phys. (N.Y.) {\bf 1982}. Vol. \underline{140}. p. 372.
\bibitem{Jackiw}
{\it Jackiw~R., Templeton~S.}
Phys. Rev. {\bf 1981}. Vol. \underline{D23}. p. 2291.
\bibitem{STZB}
{\it Sokolov~A.A. et al.}
Kalibrovochnye polya (Gauge Fields), Moscow: Moscow University Press, 1986,
[in Russian].
\bibitem{ren}
{\it Faddeev~L.D., Slavnov~A.A.}
Gauge Fields: Introduction to Quantum Theory, Benjamin-Cummings Publishing
Co., 1980.
{\it Trottier~H.D.}
Phys. Rev. {\bf 1991}. Vol. \underline{D44}. p. 464.
\bibitem{Sren}
{\it Templeton~S.}
Phys. Rev. {\bf 1981}. Vol. \underline{D24}. p. 3134.
\bibitem{Spin}
{\it Deser~S., Jackiw~R., Templeton~S.}
Phys. Rev. Lett. {\bf 1982}. Vol. \underline{48}. p. 975.
\end{thebibliography}
\end{document}